# Simulation of non-inductive vector control of permanent magnet synchronous motor based on sliding mode observer

Caiyue Zhang#
School of Electronic Engineering and Automation
Guilin University of Electronic Technology
Guilin, Guangxi Zhuang Autonomous Region

Zipin Liu#
School of Electrical and Control Engineering
Xi'an University of Science and Technology
Xi'an, Shaanxi

Bowen Xu#,*
School of Mechanical Engineering and Automation
Harbin Institute of Technology, Shenzhen
Shenzhen, Guangdong
200320109@stu.hit.edu.cn

#These authors contributed equally.

**Abstract—Permanent magnet synchronous motors (PMSM) are widely used due to their numerous benefits. It is critical to get rotor position and speed information in order to operate the motor accurately. Sensorless control techniques have emerged as a popular study area both at home and overseas. The sliding mode observer (SMO) may indirectly detect rotor position and has the benefits of easy implementation and efficient algorithms. In this study, a mathematical model for sensorless control of a PMSM is developed using SMO, vector control, and other techniques. With a surface-mounted PMSM as the study object, a mathematical model for sensorless control of PMSM is developed. PMSM's sliding mode observer model is built in the matlab/simulink environment. Experiments demonstrate that the system can track the rotor position and speed of the motor precisely and fulfill the requirements of sensorless vector control of PMSM.**

## I. INTRODUCTION

At present, PMSM is extensively utilized in industries because to its simple construction, high power factor, high power density, high accuracy, high efficiency, high torque, and ease of heat dissipation and maintenance [1]. Compared with induction motors, PMSM has permanent magnets that can provide a continuous magnetic field in the air gap, and the current of the stator is solely utilized to generate torque, which makes PMSM have a higher power factor under the same output conditions[2]. Compared with the winding - rotor synchronous motor (SM), the rotor windings of PMSM do not require DC excitation, minus the brush and sliding ring, so its cost is reduced [2].

Traditional PMSM control methods mainly include open loop control method [3], vector control(VC) method [4] and DTC method. VC is not dependent on motor parameters and has good robustness[5]. It has the disadvantage of irregular operation, resulting in large pulsations in the output current waveform [6].

Accurate control of PMSM requires not only appropriate control methods, but also accurate acquisition of rotor position information [7]. In conventional PMSM control, electromagnetic sensors, photoelectric encoders, speed generators to determine the magnetic pole location and speed of the rotor, mechanical sensors like and are most frequently utilized. [8], in order to control motor torque and speed. But adding mechanical sensors will make the motor bigger and more expensive, making the system more susceptible to interference and reducing the stability . The goal of PMSM sensorless position control is to rebuild the motor's back electromotive force [9] by monitoring the three-phase AC voltage and current of the stator, and then estimate the the rotor's position. [10] and the motor speed to achieve closed-loop control [11]. The advantages of sensorless control are improved control accuracy and anti-interference ability[12]. High frequency injection method [12], model reference adaptive method [13], extended Kalman filter method [14], intelligent algorithm [15], and SMO method [16] are the primary sensorless control techniques.

In recent years, the advantages of SMO based sliding mode control (SMC) [17] have gradually become prominent. The SMO method obtains an estimate of the internal state of a particular system [19] by measuring only the inputs and outputs of the actual system [18], and can replicate the disturbance to achieve complete compensation of the disturbance. Thanks to the above characteristics, SMO has good transient performance, fast dynamic response, insensitivity to system parameter changes and external interference, and strong robustness [20]. Is an easy to implement and commonly used robust control strategy, and can effectively target nonlinear systems with interference [21].

The research content is carried out with the acquisition of rotor position as the core. In order to determine the rotational speed, the model reference adaptive system approach suggested in reference [22] is based on stability theory.The Lyapunov equation and the Popov superstability theory ensure the asymptotic convergence of state and speed. According to reference [23], the electromagnetic torque was calculated using a stator current observer with rotor back electromotive force adaptive for the pole impact of permanent magnet brushless DC motors that stands out. Literature [24] creates a composite control approach that combines sliding mode reference adaptive control (MRAC) with an extended state observer (ESO) to view the whole disturbance.. In literature [25], full-dimensional observer is used to calculate load disturbance and estimate location error. Experimental data indicate that this approach performs well at low speeds, although the technique is challenging to implement [26].

In this paper, to address the chattering issue, a customized sliding mode plane, a high-order sliding mode, or a fuzzy control and filter combination are used of the sliding model method in the sliding mode method [27], which makes up for the issue that the traditional SMO's vibrations make it difficult to achieve the recommended controller's required performance. To achieve the speed control system's smooth transition from the independent control mode to the automatic mode, the sliding mode controller's speed following feature is used [28].

## II. SMO SIMULATION MODEL BUILDING

### A. Model Building

In this study, a SMO based on rotor position estimation of arctangent function is used, and Figure 1 displays a block schematic of it in its entirety.Clark transformation and Park transformation are used. Clark transform is the transformation of the current ( $i_a$、 $i_b$、 $i_c$ ) voltage ( $u_a$、 $u_b$、 $u_c$ ) under the natural coordinate system to the current ( $i_\alpha$、 $i_\beta$ ) or voltage ( $u_\alpha$、 $u_\beta$ ) under the stationary coordinate system. Park transformation is the transformation of the current or voltage under the stationary coordinate system to the current ( $i_d$、 $i_q$ ) or voltage ( $u_d$、 $u_q$ ) under the synchronous rotation coordinate system. Clark inverse and Park inverse are also used.

Before establishing the mathematical model, the following assumptions are made: the core magnetic saturation of the motor is negligible; eddy as well as hysteresis losses are often small, and the motor's three-phase current is a symmetrical sine wave current.

At present, most traditional SMO algorithms are designed based on the stationary coordinate system (*α-β*), so The motor's voltage equation can be revised:

$$\begin{bmatrix} u_\alpha \\ u_\beta \end{bmatrix} = \begin{bmatrix} R + \frac{dL_d}{dt} & \omega_e (L_d - L_q) \\ -\omega_e (L_d - L_q) & R + \frac{dL_q}{dt} \end{bmatrix} \begin{bmatrix} i_\alpha \\ i_\beta \end{bmatrix} + \begin{bmatrix} e_\alpha \\ e_\beta \end{bmatrix} \quad (1)$$

Where, $L_d$、 $L_q$ is the inductance of the stator; $\omega_e$ is the electric angular velocity of the stator; $u_\alpha$ and $u_\beta$ is the voltage of the stator; $i_\alpha$ and $i_\beta$ is the current of the stator; $e_\alpha$ and $e_\beta$ is the extended back electromotive force (EMF), which satisfies the following equation:

$$\begin{bmatrix} e_\alpha \\ e_\beta \end{bmatrix} = \left[ (L_d - L_q)\left( \omega_e i_d - \frac{di_q}{dt} \right) + \omega_e \psi_f \right] \begin{bmatrix} -\sin\theta_e \\ \cos\theta_e \end{bmatrix} \quad (2)$$

Where, $\psi_f$ is the magnetic chain of a permanent magnet.

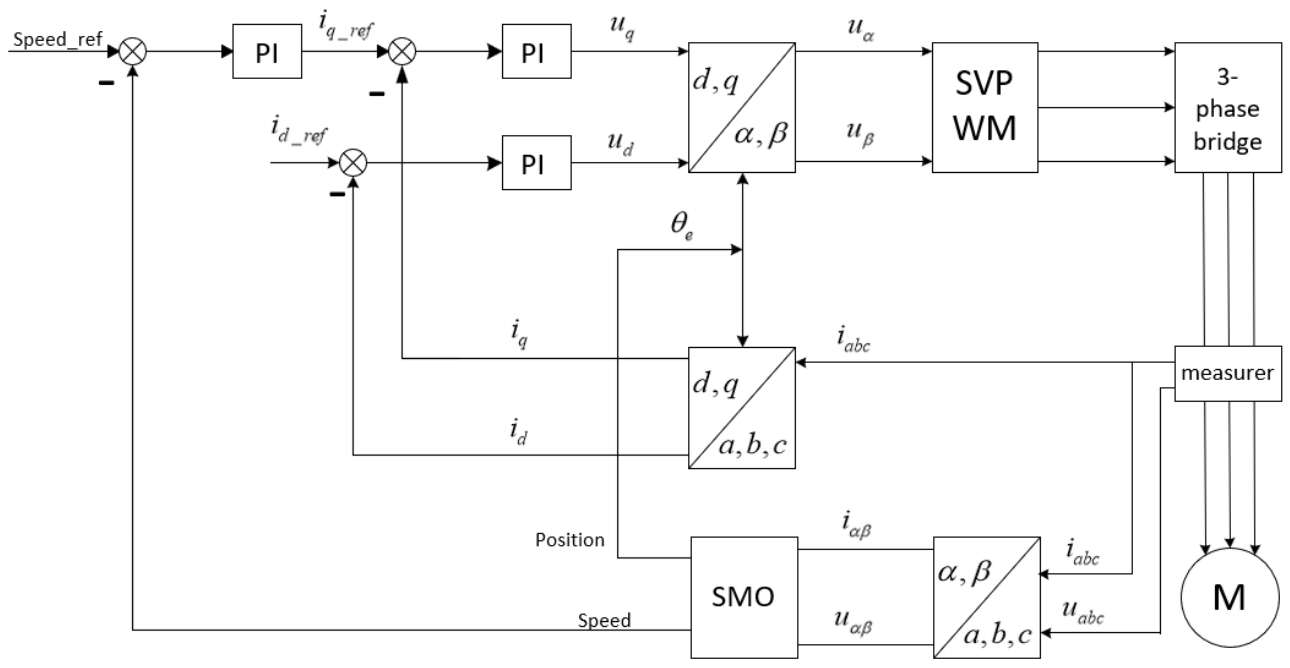

Fig. 1. Overall control block diagram based on SMO

### B. Three-phase voltage-source inverter model

The SVPWM technology not only improves the voltage utilization rate of the voltage-type inverter and the motor's capability for dynamic reaction, but also reduces the current harmonic wave and torque pulsation of the motor [29]. Therefore, the SVPWM algorithm is used as the voltage inverter control algorithm in this paper. Its theoretical basis is the principle of average equivalence, which properly combines the basic voltage vectors in a switching cycle T, so that the average voltage in a cycle is equal to the desired voltage vector. The purpose is to control the direction of the magnetic chain vector by adjusting the voltage vector's direction, and subsequently the rotor's rotation speed and orientation. There are six basic voltage vectors in a typical two-level three-phase voltage source inverter (001, 010, 011, 100, 101, 110) and two zero vectors (000, 111), respectively denoted as V1 , V2 , V3 , V4 , V5 , V6 , V0 , V7 .The six basic voltage vectors divide a 360° circular space vector into six equal sectors, each of which is 60°, and are denoated as sectors I to VI. So as long as the given voltage vector is within the output range, it can be synthesized by two adjacent voltage vectors, that is to say, the synthesis of the given vector can be completed by controlling the turn-on time of the corresponding switching tube [30]. Of course, in order to minimize the switching loss, we use the seven-section SVPWM algorithm for control.Before the control, we also need to know the number of sectors. However, knowing this is not enough, we also need to calculate the working time of each switch tube and the switching time point [30].Once you've done that, you can start taking control.

### C. Solve for Position and Velocity

As can be seen, the extended reverse electric potential contains information about the orientation and speed of the motor rotor.So we only need to obtain the extended reverse electromotive force accurately to calculate the required information. In order to make more use of the SMO to monitor the extended reverse electromotive force, the equation in (1) can be rewritten as a current state equation:

$$\begin{bmatrix} \frac{di_\alpha}{dt} \\ \frac{di_\beta}{dt} \end{bmatrix} = \frac{1}{L_d} \begin{bmatrix} -R & -(L_d - L_q)\omega_e \\ (L_d - L_q)\omega_e & -R \end{bmatrix} \begin{bmatrix} i_\alpha \\ i_\beta \end{bmatrix} + \frac{1}{L_d} \begin{bmatrix} u_\alpha \\ u_\beta \end{bmatrix} - \frac{1}{L_d} \begin{bmatrix} e_\alpha \\ e_\beta \end{bmatrix} \quad (3)$$

In a conventional SMO, the formula system utilized to calculate the back-EMF is as follows:

$$\begin{bmatrix} \frac{d\hat{i}_\alpha}{dt} \\ \frac{d\hat{i}_\beta}{dt} \end{bmatrix} = \frac{1}{L_d}\begin{bmatrix} -R & -(L_d - L_q)\omega_e \\ (L_d - L_q)\omega_e & -R \end{bmatrix}\begin{bmatrix} \hat{i}_\alpha \\ \hat{i}_\beta \end{bmatrix} + \frac{1}{L_d}\begin{bmatrix} u_\alpha \\ u_\beta \end{bmatrix} - \frac{1}{L_d}\begin{bmatrix} v_\alpha \\ v_\beta \end{bmatrix} \quad (4)$$

Where, $\hat{i}_a$ 、 $\hat{i}_\beta$ is the measured stator current value; $v_a$ 、 $v_b$ is the observer's regulated voltage input.

By deducting (3) from, the error equation for stator current is found (4):

$$\begin{bmatrix} \frac{d\tilde{i}_a}{dt} \\ \frac{d\tilde{i}_b}{dt} \end{bmatrix} = \frac{1}{L_d}\begin{bmatrix} -R & -(L_d - L_q)w_e \\ (L_d - L_q)w_e & -R \end{bmatrix}\begin{bmatrix} \tilde{i}_a \\ \tilde{i}_b \end{bmatrix} - \frac{1}{L_d}\begin{bmatrix} e_a - v_a \\ e_b - v_b \end{bmatrix} \quad (5)$$

Where, $\tilde{i}_\alpha = \hat{i}_\alpha - i_\alpha$ and $\tilde{i}_\beta = \hat{i}_\beta - i_\beta$ is the current measurement inaccuracy. The sliding mode's control rule is:

$$\begin{bmatrix} v_\alpha \\ v_\beta \end{bmatrix} = \begin{bmatrix} k \cdot sgn(\hat{i}_\alpha - i_\alpha) \\ k \cdot sgn(\hat{i}_\beta - i_\beta) \end{bmatrix} \quad (6)$$

Where, $k = max\left\{-R\left|\tilde{i}_\alpha\right| + e_\alpha sgn(\tilde{i}_\alpha), -R\left|\tilde{i}_\beta\right| + e_\beta sgn(\tilde{i}_\beta)\right\}$ is the sliding mode gain. In the experiment, after several adjustments, its value is 145.

Slipform surface is reached when the observer's state variable reaches it, $\tilde{i}_\alpha = 0$ 、 $\tilde{i}_\beta = 0$ , and the observers maintains on the slipform plane..At this point the control quantity can be seen as the equivalent control quantity, namely:

$$\begin{bmatrix} e_\alpha \\ e_\beta \end{bmatrix} = \begin{bmatrix} v_\alpha \\ v_\beta \end{bmatrix}_{eq} = \begin{bmatrix} k \cdot sgn(\tilde{i}_\alpha)_{eq} \\ k \cdot sgn(\tilde{i}_\beta)_{eq} \end{bmatrix} \quad (7)$$

To get continuous extended back-EMF estimations , a low pass filter is required, specifically:

$$\begin{bmatrix} \dot{\hat{e}}_\alpha \\ \dot{\hat{e}}_\beta \end{bmatrix} = \begin{bmatrix} (-\hat{e}_\alpha + k \cdot gn(\tilde{i}_\alpha)) / \tau_0 \\ (-\hat{e}_\beta + k \cdot sgn(\tilde{i}_\beta)) / \tau_0 \end{bmatrix} \quad (8)$$

Where, $\tau_0$ is the time constant of the low-pass filter.

However, the addition of low-pass filtering affects the amplitude and phase of the estimates of the extended inEMF. To obtain a more accurate positional information of the rotor, it is necessary to obtain it by arctangent function method and adding Angle compensation, namely:

$$\hat{\theta}_{eq} = -\arctan(\hat{e}_\alpha / \hat{e}_\beta) + \arctan(\hat{\omega}_e / \omega_c) \quad (9)$$

Where, $\omega_c$ is the cutoff frequency of a low-pass filter. With a switching frequency of 10kHz, the cutoff frequency of about 20khz is adopted, which is adjusted to 30khz in the actual experiment; $\hat{\omega}_e = \frac{\sqrt{\hat{e}_\alpha^2 + \hat{e}_\beta^2}}{\psi_f}$.

The differential operation of can be used to determine the rotational speed information (9).

*D. SMO Control Method for Sensorless PMSM*

The table-stick three-phase PMSM is studied in this study along with the PMSM vector control approach. Using the synchronous rotation index and the electromagnetic torque formula: $T_e = 3P\left[\psi_f i_q + (L_d - L_q)i_d i_q\right]/2$ (10)

It can be seen that when $i_d = 0$ , the electromagnetic torque is only proportional to $i_q$. As a result, adjusting $i_q$ is the sole way to regulate the amount of the electromagnetic torque. and the implementation is simple.

*E. Speed inner loop regulator setting*

In this study, speed internal loop and current internal loop regulators were applied. To make parameter computations easier, the motor equations of motion for a three-phase PMSM should be reconstructed as follows:

$$J\frac{d\omega_m}{dt} = T_e - T_L - \xi\omega_m \quad (11)$$

Where $\omega_m$ is the motor's mechanical angular velocity; $J$ is the moment of inertia; $\xi$ is damping coefficient; $T_L$ is the load torque.

The "active damping" principle, which is used to construct the velocity loop PI regulator's parameters, is defined as:

$$i_q = i_q^{'} - \xi_a\omega_m \quad (12)$$

When the motor is assumed to be started under no-load condition, since the $i_d = 0$ control mode is adopted, it is accessible from (10) to (12) as follows:

$$\frac{d\omega_m}{dt} = \frac{1.5p_n\psi_f}{J}i_q^{'} - \frac{1.5p_n\psi_f}{J}\xi_a\omega_m - \frac{\xi}{J}\omega_m \quad (13)$$

To meet bandwidth requirements β, the poles need to be shifted so that a transfer function of velocity with respect to Q-axis current could well be acquired, as follows：

$$\omega_m(s) = \frac{1.5p_n\psi_f}{J(s+\beta)}i_q^{'}(s) \quad (14)$$

By applying Laplace transform to (13) and comparing it with (14), the active power damping coefficient $\xi_a$ can be obtained:

$$\xi_a = \frac{J\beta - \xi}{1.5p_n\psi_f} \quad (15)$$

Since the traditional PI regulator is used in this paper, the controller expression of the velocity loop is as follows:

$$i_q^* = (K_{pw} + \frac{K_{iw}}{s})(\omega_m^* - \omega_m) - \xi_a\omega_m \quad (16)$$

Thus, the formula below can be used to obtain the PI regulator's parameters:

$$\begin{cases} K_{pw} = \frac{J\beta}{1.5p_n\psi_f} \\ K_{iw} = \beta K_{pw} \end{cases} \quad (17)$$

After plugging in the motor parameters and experimental debugging, it is found that the most basic deviation pi adjustment can be used.The value of $\beta$ is 500, $\xi_a$ is 0, $K_{pw}$ =0.004, $K_{iw}$ =2.

*F. Current inner loop regulator setting*

Rewrite the present equation under the $d-q$ coordinates to make controller design easier:

$$\begin{cases} \frac{di_d}{dt} = \frac{1}{L_d}(-Ri_d + L_q\omega_e i_q + u_d) \\ \frac{di_q}{dt} = -\frac{1}{L_d}[Ri_d + \omega_e(L_d i_d + \psi_f) - u_q] \end{cases} \quad (18)$$

As can be seen from (18), stator currents $i_d$ 、 $i_q$ generate cross-coupled electromotive forces in axis q and d respectively, which need to be decouped:

$$\begin{cases} u_{d0} = u_d + \omega_e L_q i_q = Ri_d + L_d \frac{di_d}{dt} \\ u_{q0} = u_q - \omega_e(L_d i_d + \psi_f) = Ri_q + L_q \frac{di_q}{dt} \end{cases} \quad (19)$$

Where, $u_{d0}$ 、 $u_{q0}$ represents the voltage of the axes d and q following current decoupling.

Apply the Laplace transform to (19) :

$$I(s) = G(s)U(s) \quad (20)$$

Where,

$$I(s) = \begin{bmatrix} i_d(s) \\ i_q(s) \end{bmatrix}, G(s) = \begin{bmatrix} R + sL_d & 0 \\ 0 & R + sL_q \end{bmatrix}^{-1}, U(s) = \begin{bmatrix} u_{d0}(s) \\ u_{q0}(s) \end{bmatrix}$$

In this paper, a conventional current in-loop regulator is combined with a feed-forward decoupling control strategy for obtaining the d-q axis voltage:

$$\begin{cases} u_d^* = (K_{pd} + \frac{K_{id}}{s})(i_d^* - i_d) - \omega_e L_q i_q \\ u_q^* = (K_{pq} + \frac{K_{iq}}{s})(i_q^* - i_q) - \omega_e(L_d i_d + \psi_f) \end{cases} \quad (21)$$

Where $K_{pd}$ and $K_{pq}$ represents current internal loop regulator's proportionate gain, accordingly. $K_{id}$ and $K_{id}$ represents the integral gain of current internal loop regulator's , accordingly.

If feed-forward decoupling control techniques are utilised, the regulator parameters within the current loop can only be made fully independent if the actual motor parameters used are the same as those predicted by the model.This is because I-order systems are frequent in automated control theory. The integrated three-phase PMSM's convex pole effect is also taken into consideration and the effect of the model error on the system cannot be ignored. Therefore, in order to make the designed motor model more generalisable, a tabulated control strategy is chosen for the parameter design. It has the advantages of poor model accuracy and unresponsiveness to changes in the parameters, basic design, a single parameter, and straightforward computation.

In order to calculate better, the internal model control block diagram needs equivalent transformation.As shown in Fig. 2:

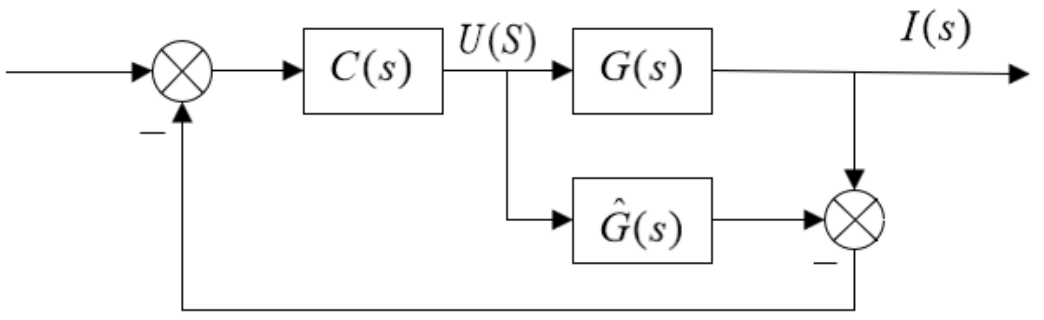

(a) Diagram of internal model control blocks

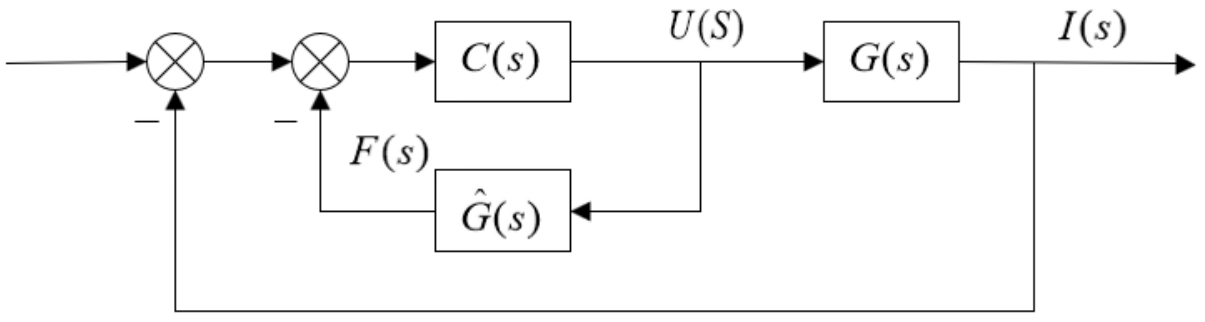

(b) Equivalent block diagram

Fig. 2. Equivalent transformation block diagram for internal mode control

The equivalent controller can be obtained:

$$F(s) = \left[I - C(s)\hat{G}(S)\right]^{-1} C(s) \quad (22)$$

Where $I$ is the identity matrix.

When $\hat{G}(s) = G(s)$ , if the system's feedback connection is absent, the system transfer function is as follows:

$$G_c = C(s)G(s) \quad (23)$$

As a result, the system is only stable if and when $C(s)$ and $G(s)$ are stable.

According to $\hat{G}(s) = G(s)$ , the current loop of the control system may be classified as a first-order process since the motor's electromagnetic time constant is much less than its mechanical time constant, the definition is as follows:

$$C(s) = \hat{G}^{-1}(s)L(s) = G^{-1}(s)L(s) \quad (24)$$

Where, $L(s) = \frac{aI}{s+a}$ , $a$ is the design parameter, and its calculation formula is $\frac{2\pi R}{L}$ , $L$ is the inductance, $R$ is the resistance.

By substituting (24) into (22), you may get the internal model controller.:

$$F(s) = a\begin{bmatrix} L_d + \frac{R}{s} & 0 \\ 0 & L_q + \frac{R}{s} \end{bmatrix} \quad (25)$$

By substituting (25) into (23), we can get:

$$G_c(s) = \frac{a}{s+a}I \quad (26)$$

By comparing (26) with (21), we can get:

$$\begin{cases} K_{pd} = aL_d \\ K_{id} = aR \\ K_{pq} = aL_q \\ K_{iq} = aR \end{cases} \quad (27)$$

The final values obtained by substituting the motor parameters and adjusting them during the experiment are: $K_{pd} = K_{pq}$ =120.54, $K_{id} = K_{iq}$ =70440.

The simulation model is built according to (21) and (27), as shown in Figure 3.Among them, the continuous PI regulator is used for simulation modeling.At this point, the parameter setting of PI regulator is complete.

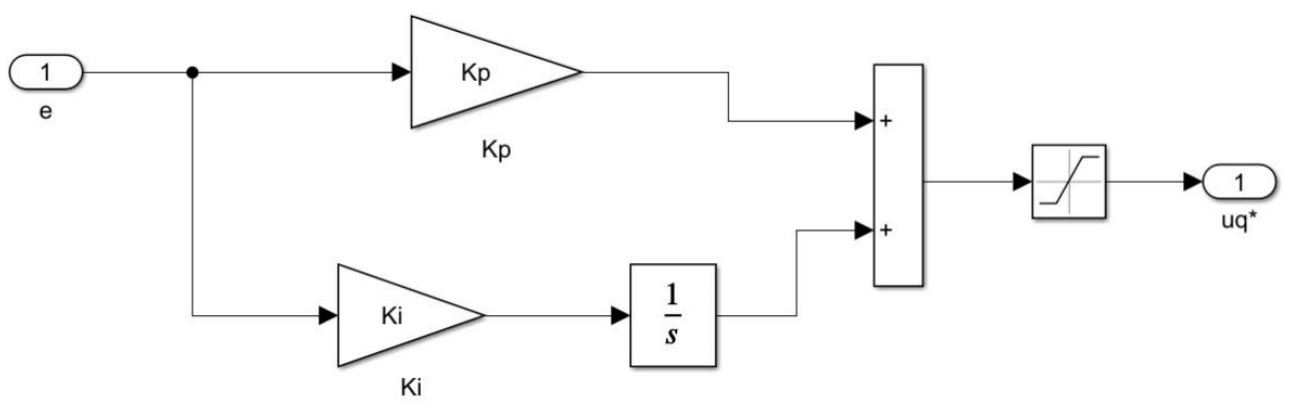

Fig. 3. Block diagram of the closed loop regulation controller

Because in the process of simulation parameters adjustment, found that the approximate control effect can be achieved even if the " $\omega_e L_q i_q$ " and " $\omega_e (L_d i_d + \psi_f)$ " in (21) are omitted, the type changed (21) into (28) :

$$\begin{cases} u_d^* = (K_{pd} + \dfrac{K_{id}}{s})(i_d^* - i_d) \\ u_q^* = (K_{pq} + \dfrac{K_{iq}}{s})(i_q^* - i_q) \end{cases} \quad (28)$$

## III. Integrated Experimental and Simulation Results

An genuine permanent magnet synchronous motor (60ST-M00630) was chosen for simulation in this research in order to confirm the algorithm's functionality. Table 1 displays the motor's particular parameters:

Tbl. 1. The main motor settings utilized in the experiment

| parameter | The numerical |
|---|---|
| Rated voltage U/V | 220 |
| Rated power P/W | 200 |
| Rated speed n/(r/min) | 3000 |
| Rated torque Tn/(N·M) | 0.637 |
| Rated current I/A | 1.5 |
| Rotor inertia J/(kg·m²) | $0.17 \times 10^{-4}$ |
| Permanent magnet flux$\varphi_f$ / $w_b$ | 0.3477 |
| Stator resistance (between lines)Ls/H | 11.6 |
| Stator inductance (between lines)Ls/H | 0.022 |
| Frequency of switching $f_s$ / $kHz$ | $1 \times 10^4$ |
| Polar logarithm P | 4 |

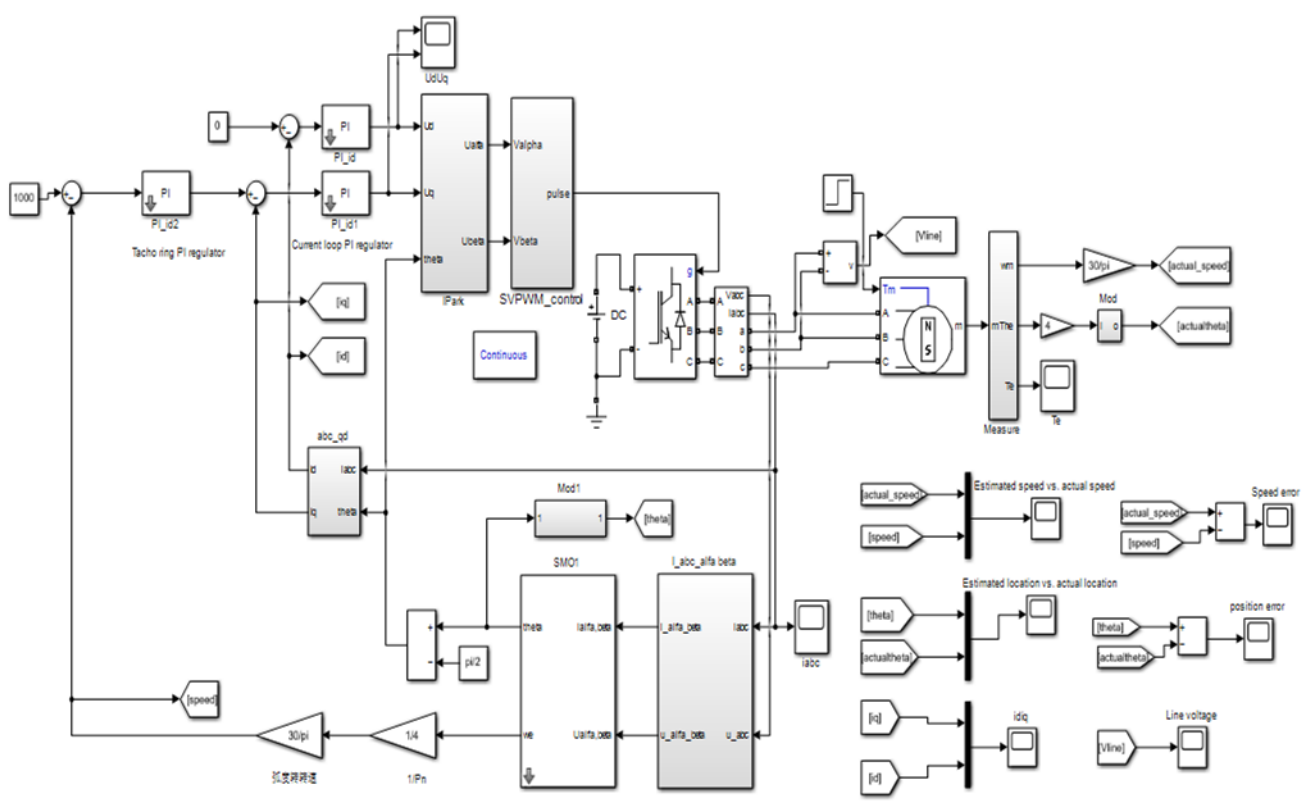

Fig. 4. PMSM vector control system simulation model based on SMO

Figure 1 depicts the mathematical simulation model of the motor. it has a simulation duration set to 0.1s, the simulation step is set as $2 \times 10^{-7}$s, and the fixed step size ode3 (Bogacki-Shampine) algorithm is selected.

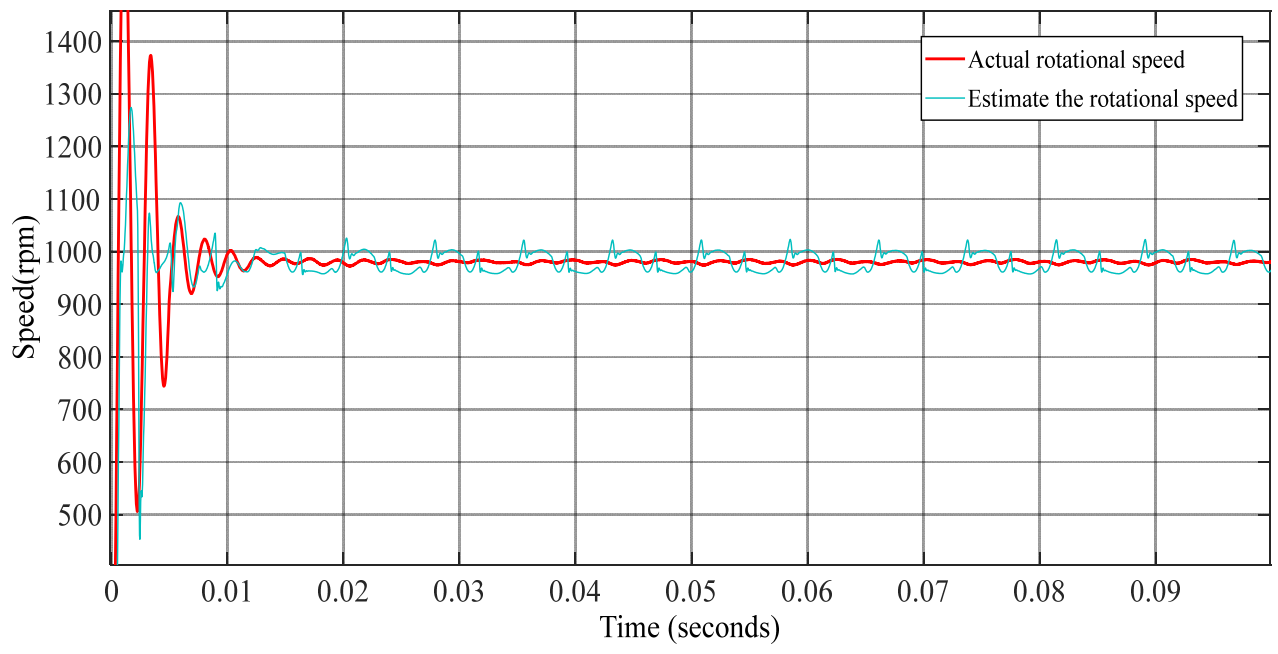

(a) Speed measurement curve of sliding mode observer

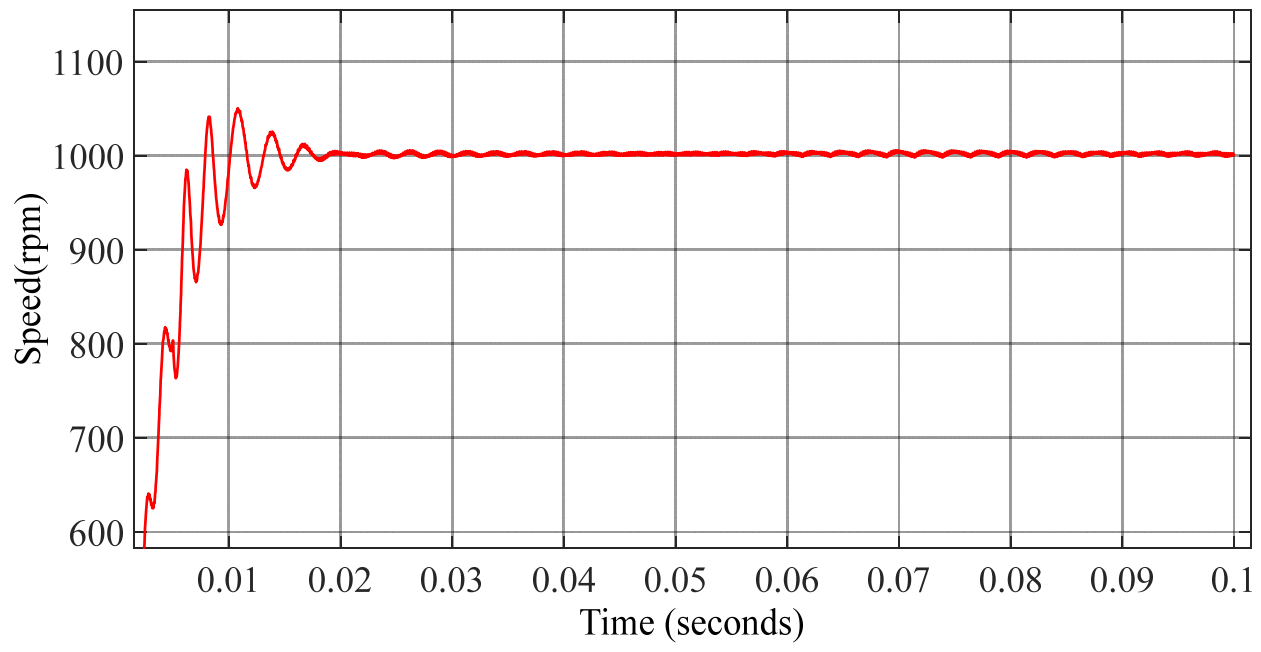

(b) Speed measurement curve of PI controller

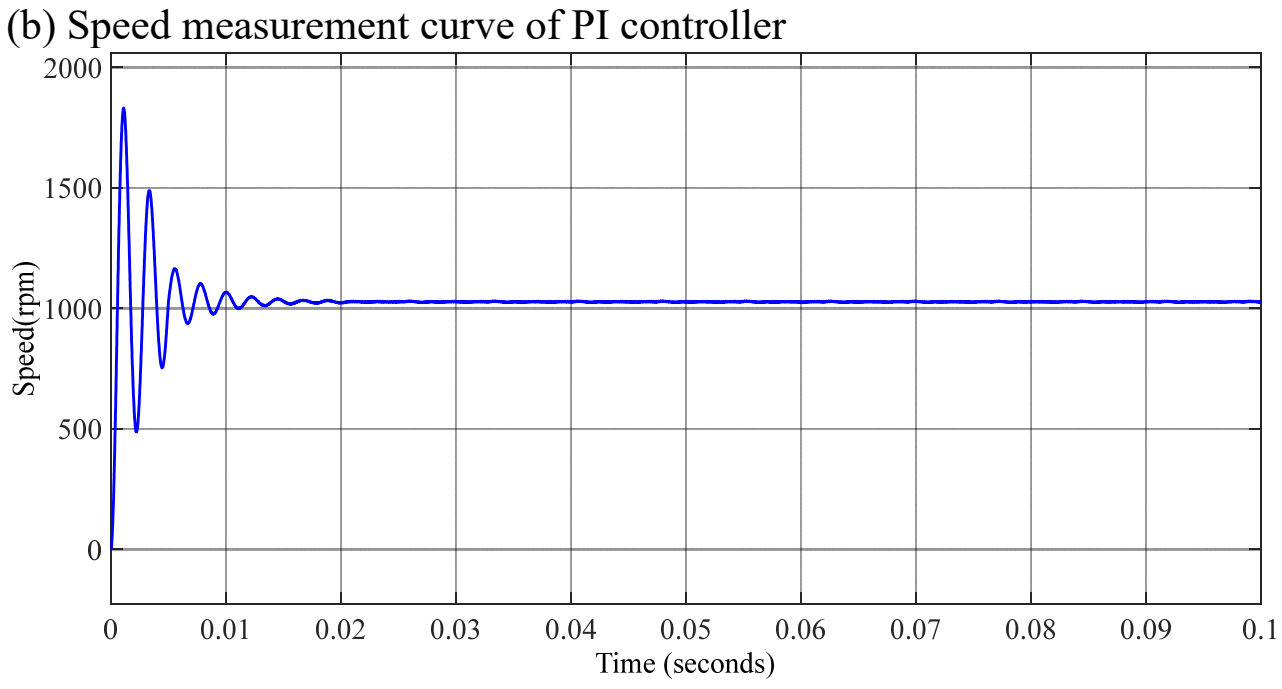

(c) Speed measurement curve of open loop controller

Fig. 5. Comparison of speed curves based on different controllers

Figure 5. (a) shows the closed-loop control effect curve using a SMO. FIG. b shows the PI closed-loop control effect when the position sensor is used for speed measurement; FIG.

c shows the open-loop control effect when the position sensor is used for speed measurement.The speed waveforms in Figures a and b show that the PI controller has significant dampening and little overshoot. After measurement, the overshoot of the SMO is 70% and the time to reach the buffer zone with a speed difference of 5% is 0.007 s. The overshoot of the PI controller is 5% and the time to reach the buffer zone with a speed difference of 5% is 0.01 s.The overshoot of the open-loop controller is 83.2% and the adjustment time is 0.011s.It can be seen that the three controllers all have varying degrees of overshoot oscillation after the motor starts for a period of time. Combining the two parameters of overshoot and adjustment time, the synovial observer has shorter adjustment time and faster speed response than the other two controllers.

When the speed is set as 1000r/min, 800r/min and 600r/min respectively, the waveform changes of the motor at different speed values are compared and analyzed.

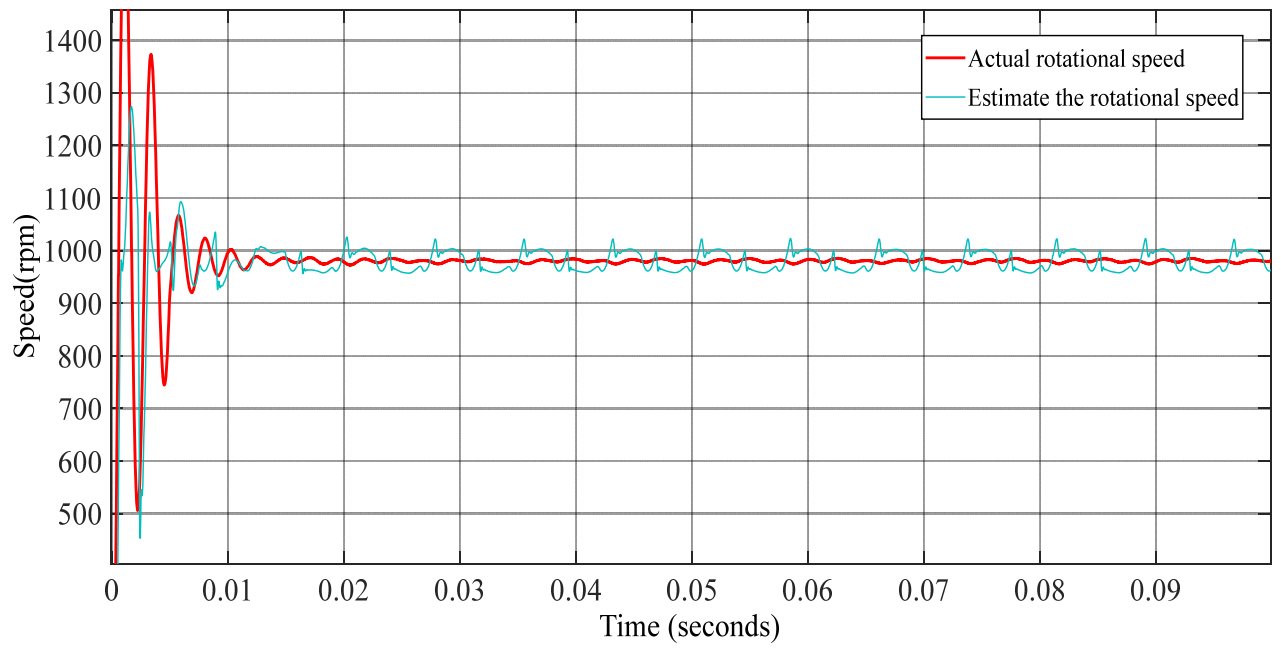

(a)Speed profile at 1000 revolutions per minute

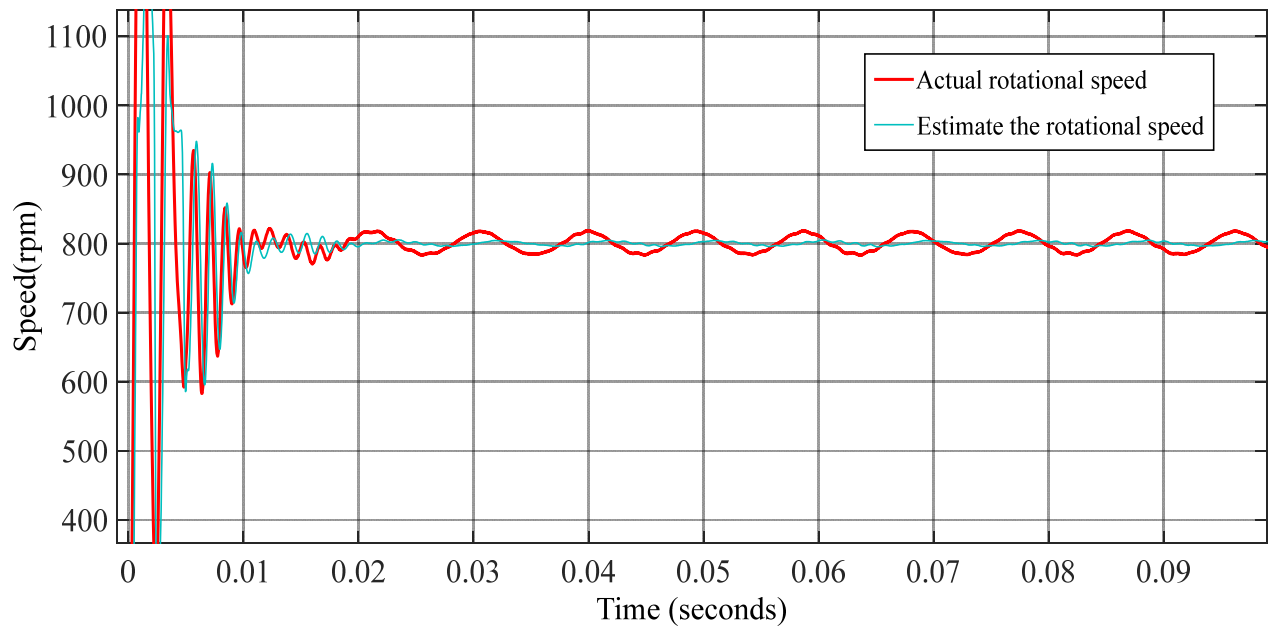

(b) Speed profile at 800 revolutions per minute

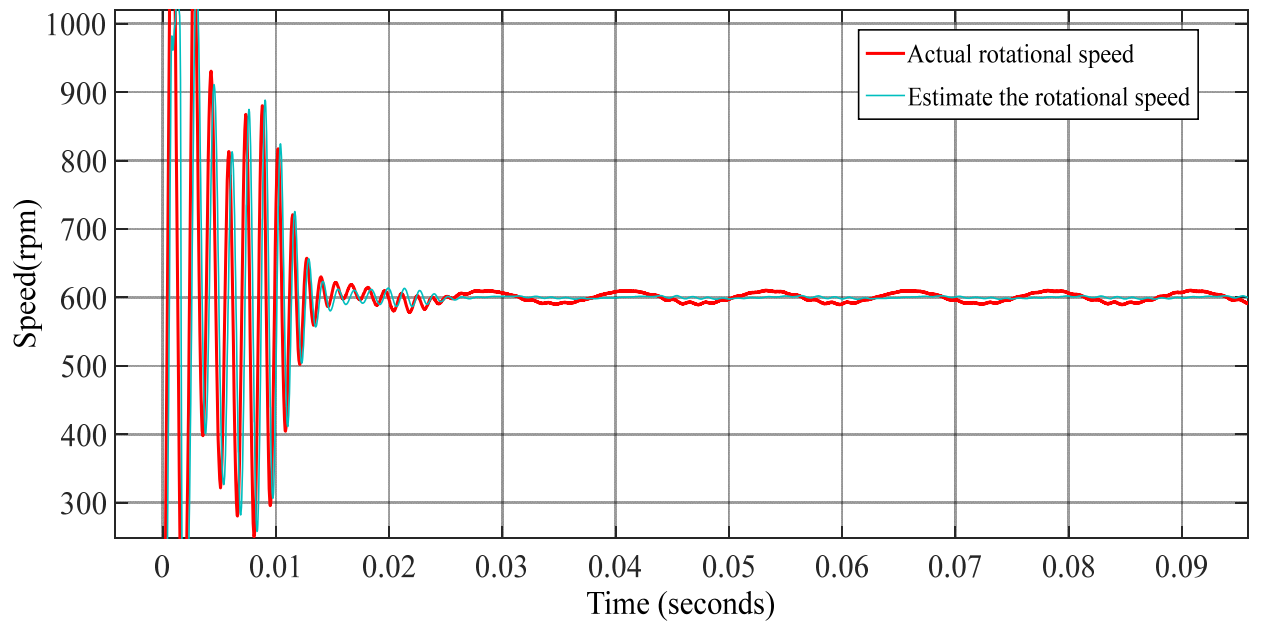

(c) Speed profile at 600 revolutions per minute

Fig. 6. Corresponding speed curves when different speed values are set for the motor

From the comparative graphs corresponding to different speed response curves when the motor is set to three different speed values in Figure 6, as can be observed, the greater the speed preset value, the smoother the output and the smaller the oscillation amplitude.

To further evaluate the synchronous motor's permanent magnet SMO's performance under control, the initial speed setting was 800r/min, the speed value was set to 600r/min at t=0.06s, the speed value was set to 1000r/min at 0.12s and the time for the system simulation was set to 0.2s. The voltage, current, motor torque and speed following in the control system were observed.

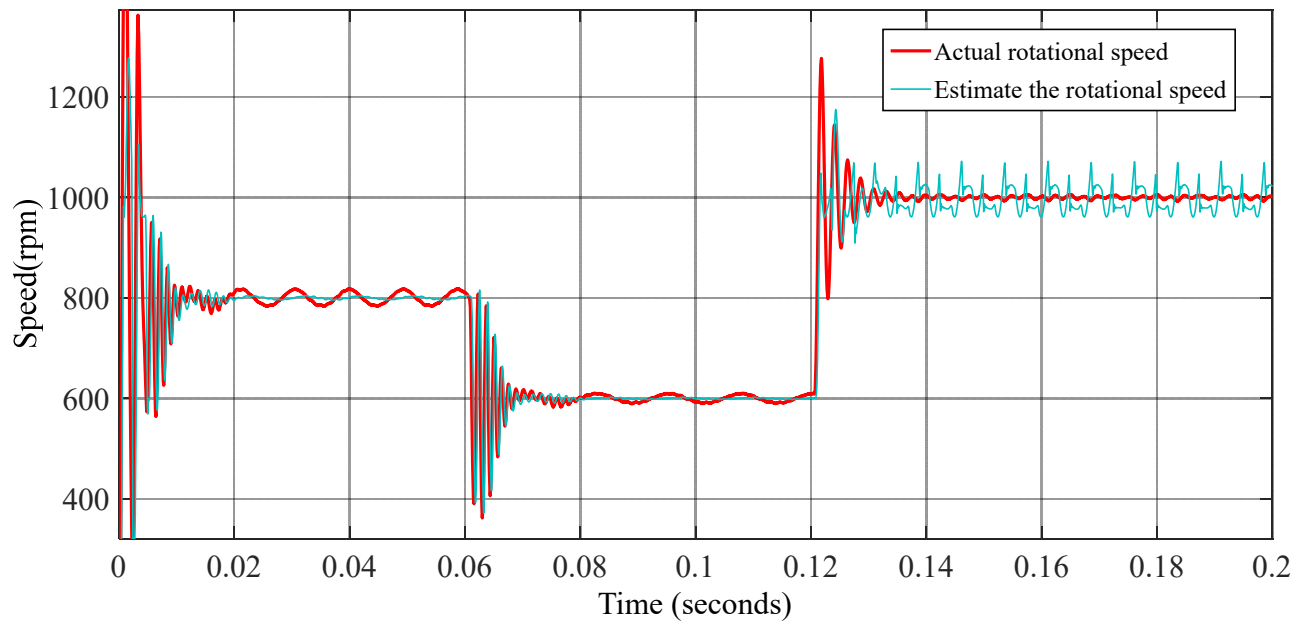

Fig. 7. Speed variation curve while the engine is accelerating and decelerating

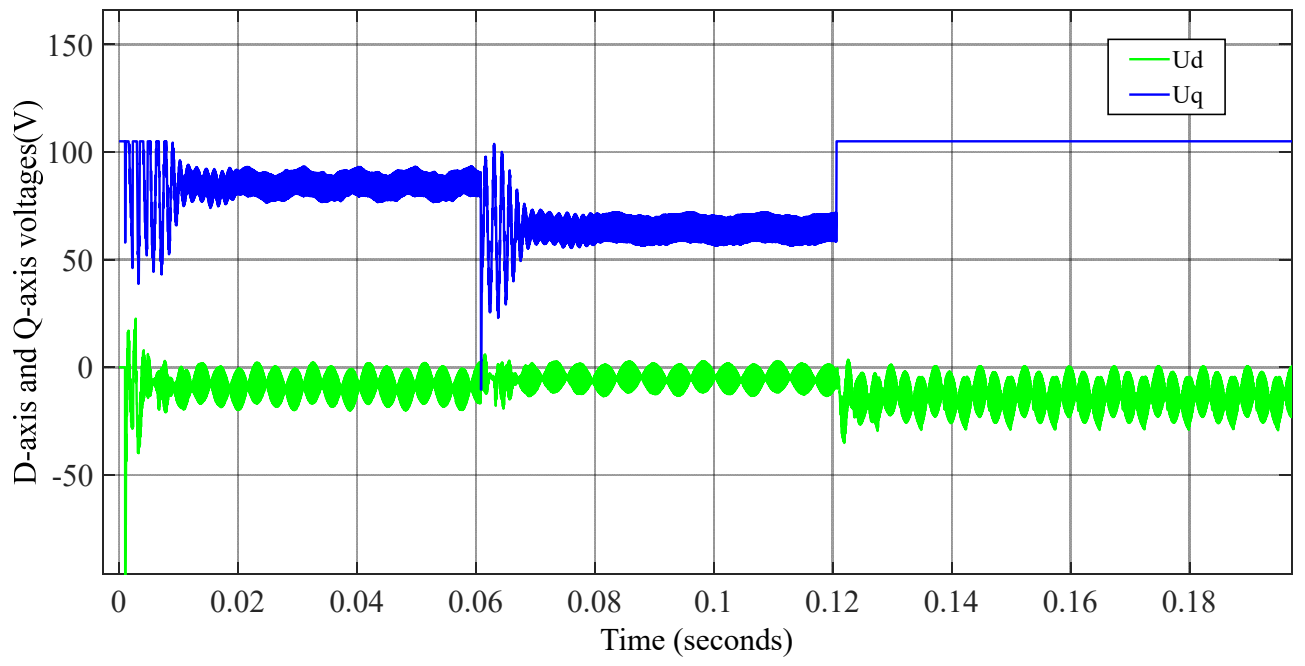

Fig. 8. D-axis and q-axis curves voltage changes while the engine is accelerating and decelerating

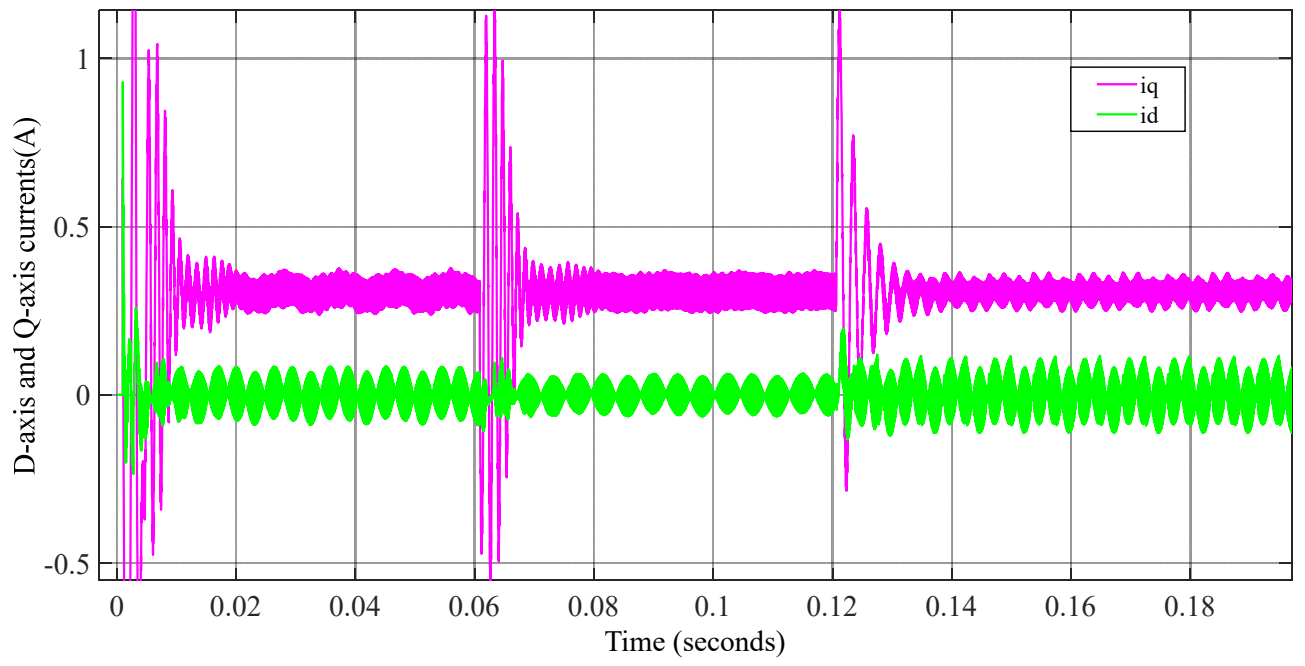

Fig. 9. D-axis and q-axis curves current variation while the engine is accelerating and decelerating

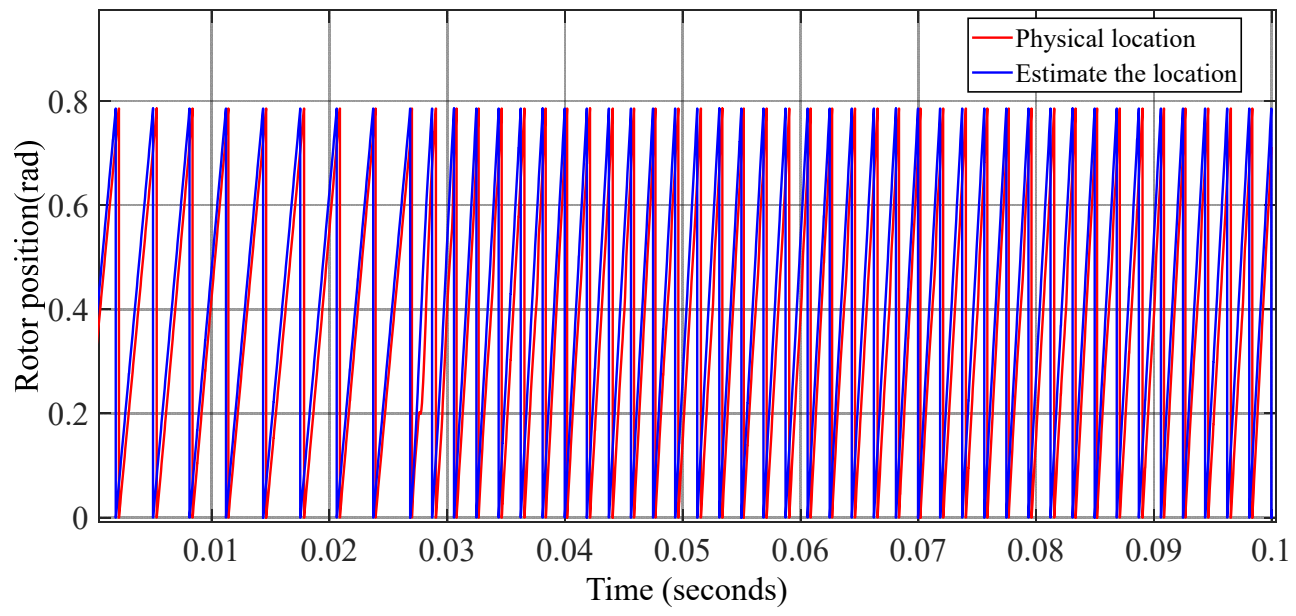

Fig. 10. Position following curve while the engine is accelerating and decelerating

Figure 7 shows that the motor speed stabilises at a predetermined value after a short oscillation when the speed is changed for a certain period of time. Figure 8 shows that the q-axis voltage increases and decreases with increasing and decreasing speed. Figure 9 shows that the q-axis current oscillates at different frequencies as the speed is varied. Figure 10 shows that the position estimation algorithm is still able to perform good position following when the motor is being accelerated or decelerated at any given moment, the small following delay in the figure is due to a flaw in the slide film algorithm itself.

The stability of the system was further tested by sudden loading. When t=0.035s, the rated torque of 0.637 was input and the variation curve of motor speed was observed, as shown in Fig. 11.

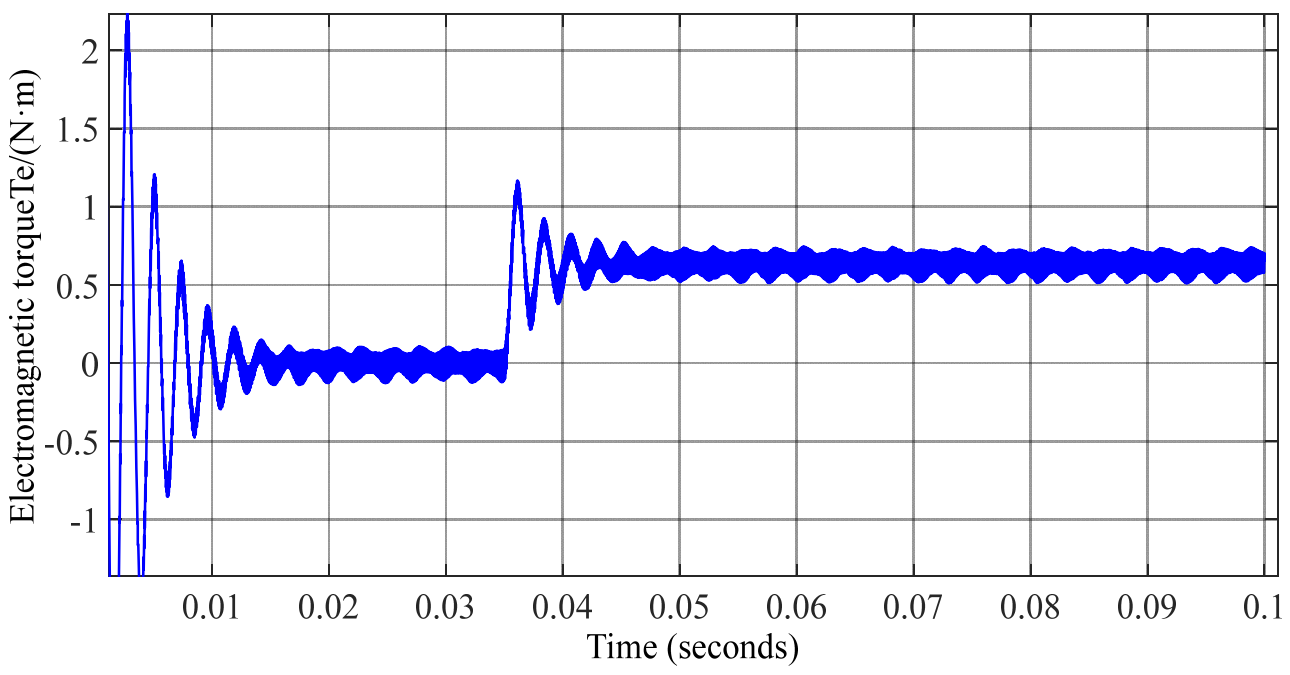

Fig. 11. Increase in rated torque of 0.637 at t=0.035s

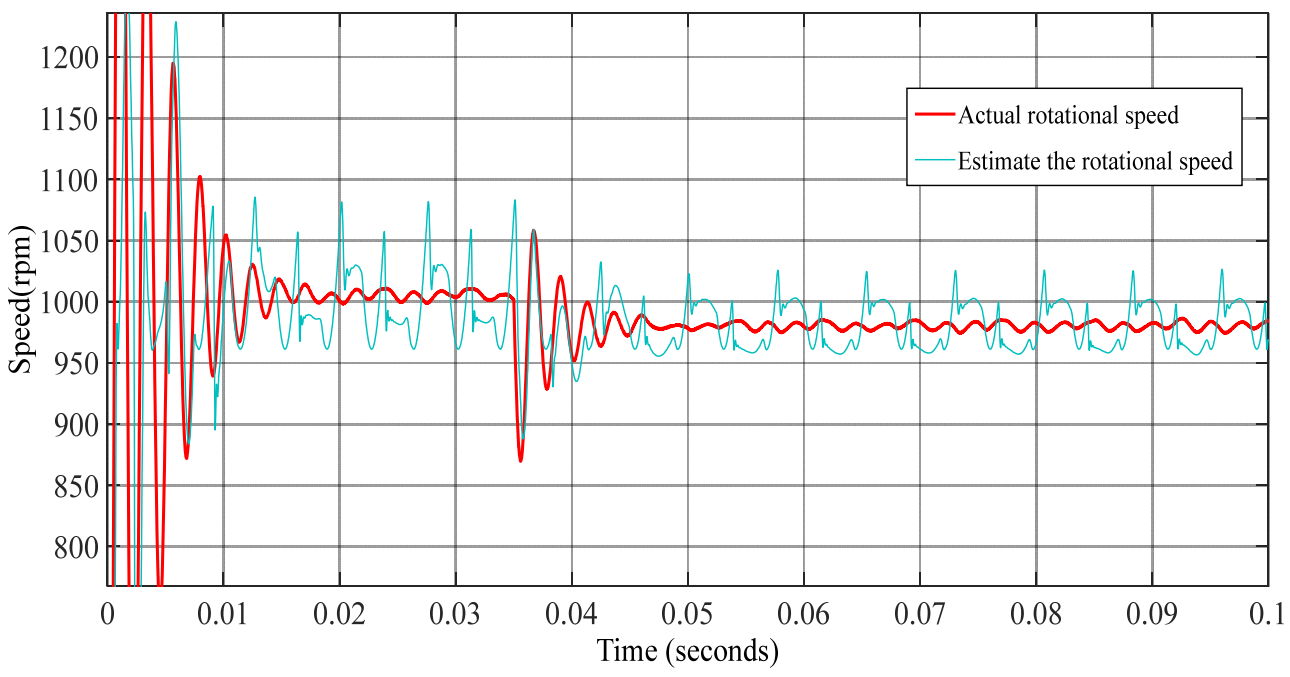

Fig. 12. Motor speed curve during sudden load application

As can be seen in Figure 12, after adding 0.637 rated torque at t=0.035s, the motor speed first oscillates slightly and then decelerates slightly, and after adjustment the speed is eventually stabilised around a predetermined value.

## IV. Conclusion

Conclusion: The PMSM sensorless control model presented in this research is based on the SMO, and for verification, in the matlab/simulink environment, the control simulation model is created. According to the experimental findings, when compared to open loop and PI controllers with position sensors, SMO has faster response speed and better dynamic performance. When set to different speed values, the speed stabilises around the preset value after a short oscillation, and for acceleration and deceleration, the predicted position can correspond to the rotor's real location. It has a certain load capacity and strong stability. It can be shown that SMO based position sensorless algorithm is less affected by system parameters and external interference, strong anti-interference ability, and easy to implement. By selecting appropriate parameters, the actual motor control performance criteria can be satisfied by sensorless control technology based on SMO.

In addition, compared to other sensorless motor control techniques, the method of the slide film observer is straightforward and easy to use in real systems. Because of the high frequency jitter caused by the use of discontinuous switching functions in the slide film observer, there is always a delay in the position solution, even after the high frequency jitter has been removed with a low-pass filter. More studies are required for reducing high-frequency jitter in the slide film observer and provide reliable phase correction.